\documentclass[secnumarabic,amssymb, nobibnotes, aps, prd]{revtex4-2}

\usepackage{tocloft}
\usepackage{graphicx}
\usepackage{amsmath}
\usepackage{subcaption}
\usepackage{hyperref}
\hypersetup{colorlinks,linkcolor={red},citecolor={blue},urlcolor={red}}
\usepackage{setspace}
\usepackage{xcolor}
\pagecolor{white}

\usepackage{multirow}
\color{black}

\usepackage{graphicx}
\usepackage{amsmath}
\usepackage{subcaption}
\usepackage{hyperref}
\hypersetup{colorlinks,linkcolor={red},citecolor={blue},urlcolor={red}}
\usepackage{setspace}
\usepackage{xcolor}
\pagecolor{white}

\usepackage{multirow}

\begin{document}
	\title{Holographic Einstein Rings of AdS black holes with higher derivative corrections in presence of string cloud}
	
	\author{Abhishek Baruah$^{1,2}$}
	\email{$rs_abhishekbaruah@dibru.ac.in$}
		
	\author{Bidyut Hazarika$^1$}
	\email{$rs_bidyuthazarika@dibru.ac.in$}
	
	\author{Prabwal Phukon$^{1,3}$}
	\email{$prabwal@dibru.ac.in$}
	
	\affiliation{$^1$Department of Physics, Dibrugarh University, Dibrugarh, Assam, 786004.\\$^2$Department of Physics, Patkai Christian College, Ch\"umoukedima, Nagaland, 797103\\$^3$Theoretical Physics Division, Centre for Atmospheric Studies, Dibrugarh University, Dibrugarh, Assam, 786004.\\}
	
	\begin{abstract}
		This paper seeks to explore the holographic optical appearance of an AdS black hole with higher-derivative corrections in the presence of a string cloud, drawing on the AdS/CFT correspondence and wave optics. We introduce a Gaussian wave source that oscillates at the AdS boundary and propagates through the bulk. The resulting response function is then analyzed using an imaging system to study the optical properties of the black hole. Depending on the observer’s position and variations in system parameters, the resulting holographic image takes on different forms, including a well-defined Einstein ring, deformed luminous patterns, or isolated bright spots. Notably, when the observer is positioned at the north pole, the image consistently features a bright ring at the photon sphere, encircled by concentric stripes. Our findings reveal a strong correlation between the Einstein ring’s location and the photon sphere radius predicted by geometric optics, reinforcing the connection between wave optics and gravitational lensing. Additionally, we examine the influence of higher-derivative corrections and string cloud parameters on the observed optical features.
	\end{abstract}

	\maketitle
	\vspace{0.2cm}
	
\section{Introduction}
Black holes are among the most enigmatic entities emerging from the fabric of General Relativity (GR) and stand as one of its most profound and captivating predictions.Since its ground breaking formulation in 1915, GR has remained the bedrock of modern gravitational physics, offering an elegant framework to describe the curvature of spacetime. The historic detection of gravitational waves (GWs) from black hole mergers by the Laser Interferometer Gravitational-Wave Observatory (LIGO) provided an extraordinary validation of the existence of black holes, heralding a new era in our understanding of gravity.\cite{ligo}. This breakthrough was soon followed by the Event Horizon Telescope (EHT) capturing the first-ever image of a supermassive BH in the M87 galaxy, and later, Sagittarius A* (SgrA*), the BH at the center of the Milky Way \cite{m87a, m87b, m87c, m87d, m87e, m87f}. These images revealed a dark central shadow surrounded by a bright photon ring, whose morphology depends on the nature of the BH and the underlying gravitational theory \cite{Shadow1,Shadow2,Shadow3,Shadow4}. Studies have shown that BH shadows and photon spheres can serve as powerful tools for testing alternative gravity theories, as deviations from general relativity influence their size and shape \cite{Shadow5,Shadow6,Shadow7,Shadow8,Shadow9,Shadow10,Shadow11,Shadow12}.  

Most studies on BH imaging rely on geometric optics, which traces light rays around a BH to construct its shadow. However, an alternative approach based on wave optics has recently gained attention, particularly within the framework of the AdS/CFT correspondence \cite{AdSCFT1,AdSCFT2,AdSCFT3}. The AdS/CFT correspondence suggests a duality between a gravitational system in an anti-de Sitter (AdS) spacetime and a conformal field theory (CFT) on its boundary \cite{AdSCFT1,AdSCFT2}. This duality has been widely applied to various strongly coupled systems, including quantum chromodynamics and condensed matter physics \cite{HoloQM1,HoloQM2,HoloQM3,HoloQM4,HoloQM5,HoloQM6,HoloQM7,HoloQM8,HoloQM9,
HoloQM10,HoloQM11,HoloQM12}.  

A pioneering study by Hashimoto et al. introduced a wave-optics-based method for constructing holographic images of BHs \cite{prl,Hashimoto2020}. They considered a  CFT on a two-dimensional sphere  \(S^2\) at finite temperature and introduced a localized, time-dependent Gaussian wave source on the AdS boundary. The response function of the CFT was then mapped to a holographic image, revealing a structure resembling an Einstein ring.  Our focus is on the one-point function of a scalar operator \(\mathcal{Q}\), which is influenced by a localized Gaussian source \(J_{\mathcal{Q}}\) oscillating at frequency \(\omega\). The local response function we study takes the form \(e^{-i\omega t} \langle \mathcal{Q}(\vec{x}) \rangle\). The time-dependent source \(J_{\mathcal{Q}}\) acts as a boundary condition for the bulk scalar field, effectively injecting waves into the bulk from the AdS boundary. These waves then propagate through the black hole spacetime, eventually reaching different points on the boundary \(S^2\). The amplitude of the scalar field at these points corresponds to the expectation value \(\langle \mathcal{Q}(\vec{x}) \rangle\) in the boundary QFT. Using concepts from wave optics, we derive a mathematical transformation that converts the response function \(\langle \mathcal{Q}(\vec{x}) \rangle\) into an image of the black hole on a virtual screen. The reconstructed image, denoted as \(|\Psi_{Sc}(\vec{x}_{Sc})|^2\), is given by:

\[
\Psi_{Sc}(\vec{x}_{Sc}) = \int_{|\vec{x}|<d} d^2x \, \langle \mathcal{O}(\vec{x}) \rangle e^{-i\omega f \vec{x} \cdot \vec{x}_{Sc}}
\]

where \(\vec{x} = (x, y)\) and \(\vec{x}_S = (x_{Sc}, y_{Sc})\) represent Cartesian-like coordinates on the boundary sphere \(S^2\) and the virtual screen, respectively. The origin of these coordinates is set at the observation point. This process is essentially a localized Fourier transform of the response function over a small patch of radius \(d\), incorporating a suitable window function. The function \(f\) determines the magnification of the image on the screen. From an optical perspective, this transformation mimics a lens with focal length \(f\) and radius \(d\), reconstructing the black hole’s visual representation.
This technique demonstrated that within a well defined  QFT, a dual BH could produce an Einstein ring, offering a direct method to probe gravitational duality. Subsequent studies extended this framework to various modified gravity backgrounds, exploring how different parameters influence the Einstein ring's size and brightness \cite{holo1,holo2,holo3,holo4,holo5,holo6,holo7,holo8,holo9,holo10,holo11,holo12}.  \\

The study of strongly coupled gauge theories is challenging due to the absence of systematic computational approaches. The AdS/CFT correspondence provides a valuable framework by relating a strongly coupled $\text{SU}(N_c), \mathcal{N} = 4$ Super Yang-Mills (SYM) theory at the boundary to a weakly coupled gravity theory in a higher-dimensional AdS spacetime. Within this duality, black hole solutions in the bulk correspond to the deconfined phase of the boundary gauge theory, while the confined phase is associated with a pure AdS background.  A more realistic extension of this framework involves the inclusion of fundamental quarks and baryons, which are absent in the standard $\mathcal{N} = 4$ SYM theory. This is achieved by introducing string clouds, where the endpoints of hanging strings at the boundary represent quark-antiquark pairs, while the strings themselves correspond to gluonic interactions. In this work, we investigate black hole solutions in Gauss-Bonnet gravity in the presence of string clouds, which alter the phase structure of these solutions. The string clouds introduce fundamental strings extending from the boundary to either the black hole horizon or the spacetime center, while the Einstein-Gauss-Bonnet (EGB) framework accounts for higher-order curvature corrections. The higher-order curvature corrections in this theory can be understood as subleading contributions arising from the 't Hooft coupling.  The gravitational action in five-dimensional Gauss-Bonnet gravity is given by  \cite{main}

\begin{equation}
S = \frac{1}{16\pi G} \int d^5x \sqrt{-g} \left( R - 2\Lambda + \alpha L_{\text{GB}} \right) + S_m,
\label{action}
\end{equation}

where $G$ is the  gravitational constant, \( R \) represents the Ricci scalar, and \( \Lambda \) is the cosmological constant. The metric tensor of the spacetime is denoted by \( g_{\mu\nu} \), and \( g \) represents its determinant. The Gauss-Bonnet correction term, arising from quantum field renormalization, is given by   \cite{main}

\begin{equation}
L_{\text{GB}} = R^2 + R_{\mu\nu\rho\sigma} R^{\mu\nu\rho\sigma} - 4 R_{\mu\nu} R^{\mu\nu},
\end{equation}

where \( \alpha \) is the Gauss-Bonnet coupling coefficient.

The additional $S_m$ term in eq.  \eqref{action} accounts for the presence of string clouds. Since this term represents the contribution of a large number of strings, it takes the form   \cite{main}

\begin{equation}
S_m = -\frac{1}{2} \sum_i T_i \int d^2\xi \sqrt{-h} h^{\beta\gamma} \partial_\beta X^\mu \partial_\gamma X^\nu g_{\mu\nu}.
\end{equation}

Here, the integral is taken over the string worldsheet, with \( h_{\beta\gamma} \) being the worldsheet metric and \( \beta, \gamma \) corresponding to the worldsheet coordinates. The quantity \( T_i \) represents the tension of the \( i \)-th string.  \\

The motivation behind this study is to explore the holographic imaging of an AdS black hole with higher-derivative corrections in the presence of a string cloud. The groundbreaking image of the supermassive black hole in M87, captured by the Event Horizon Telescope (EHT), has offered profound insights into black hole imaging and its alignment with theoretical predictions. Building on this development, we extend the investigation of gravitational duals by analyzing the response function in a thermal quantum field theory (QFT) framework for an AdS black hole with higher-derivative corrections and a surrounding string cloud. Specifically, we examine how this response function can be utilized to reconstruct the optical appearance of the dual black hole, giving rise to an Einstein ring. Using the AdS/CFT correspondence and wave optics techniques, we study the effects of these corrections on the structure of the holographic Einstein ring. Our findings reveal a strong correlation between the ring's position and the photon sphere radius predicted by geometric optics. This agreement between the Einstein ring angle observed in holographic imaging and the incident angle derived from geometric optics underscores a deep consistency between these two approaches.\\
\section{Overview of the Black Hole Solution}
The AdS black holes solution with higher derivative corrections in presence of
string cloud takes the form 

\begin{equation}
ds^2 = -N(r) dt^2 + \frac{1}{N(r)} dr^2 + r^2 g_{ij} dx^i dx^j,
\end{equation}

where \( g_{ij} \) represents the metric on the \((4-1)\)-dimensional boundary. The metric components are given by   \cite{main}

\begin{equation}
N(r) = 1 + \frac{r^2}{4\alpha} \left(1 - \sqrt{1 + \frac{32\alpha M}{r^4} - \frac{8\alpha}{l^2} + \frac{16 \beta \alpha}{3r^3}} \right),
\end{equation}
 Here, \( l \) denotes the AdS radius, which is related to the cosmological constant through \( \Lambda = -\frac{6}{l^2} \), and \( M \) represents an integration constant corresponding to the black hole mass. Again  $\alpha$ is the GB coupling constant and $\beta$ represents  the  density  of  the strings that are uniformly distributed
over the three spatial directions .The mass of the black hole can be calculated by setting $N(r=r_+)=0$, where $r_h$ is the event horizon radius. The ADM mass of the black hole can be expressed in terms of the horizon radius as:

\begin{equation}
M = \frac{3r_{h}^4 + 3l^2 r_h^2 - 2 \beta l^2 r_h + 6l^2\alpha}{12l^2}.
\end{equation}

Again, the temperature is determined as

\begin{equation}  
T = \frac{1}{4\pi} \frac{\partial N}{\partial r_h} = \frac{6r_h^3 + 3l^2 r_h - \beta l^2}{6\pi l^2 (r_h^2 + 4\alpha)}
\label{temp}
\end{equation}  

This solution remains asymptotically AdS, and its stability against decay into pure AdS is further investigated in  \cite{main}. \\
\section{Formation of Response function}

First, we convert the metric function to Eddington-Finkelstein(EF) coordinate system\cite{holo13} in order to solve  the Klein-Gordon equation which is given by \cite{prl} 

\begin{equation}
D_{a}D ^{a}\tilde{\Psi}=0  .
\end{equation}

  The coordinates in incident EF coordinate can be expressed as \cite{holo13} 
  \begin{equation}
\mathrm{d}s^{2} =\frac{1}{z^{2} }\Big[-N(z)\mathrm{d}\upsilon  ^{2}-2\mathrm{d}z \mathrm{d}\upsilon + \mathrm{d}\theta ^{2}+\sin ^{2}\theta \mathrm{d}\varphi ^{2}  \Big].
\end{equation}

The asymptotic  solution of the massless scalar field near the AdS boundary is obtained to be \cite{holo13} 
 \begin{eqnarray}
\tilde{\Psi}(\upsilon ,z,\theta, \varphi  )&=&J_{\mathcal{Q} }  (\upsilon ,\theta, \varphi ) +y\partial_{\upsilon }  J_{\mathcal{Q} } (\upsilon ,\theta, \varphi )\nonumber\\&+&
\frac{1}{2} z^{2} D^{2} J_{\mathcal{Q} }  (\upsilon ,\theta, \varphi )+\left \langle \mathcal{Q}  \right \rangle z^{3} +Q(z^{4} ),
\end{eqnarray} 
where $D^{2}$ denotes the scalar Laplacian and $J_{\mathcal{Q} } $ is the source function defined on the boundary based on the AdS/CFT framework \cite{w2} which is independent of the boundary coordinates.  $\left \langle \mathcal{Q}  \right \rangle$ is the corresponding response function in the dual CFT.We choose a Gaussian wave packet that is monochromatic and axisymmetric as the wave source$J_{\mathcal{Q} }$,  placed at the South Pole $(\theta _{0} =\pi )$ of the AdS boundary.
\begin{eqnarray}
J_{\mathcal{Q} }  (\upsilon ,\theta )&=&e^{-i\omega \upsilon }(2\pi\eta ^{2} )^{-1} \mathrm{exp}[-\frac{(\pi -\theta )^{2} }{2\eta ^{2} }] \nonumber\\
&=&e^{-i\omega \upsilon }\sum_{l=0}^{\infty }C_{l0} Y_{l0}(\theta )  ,
\end{eqnarray} 
The parameter $\eta$ characterizes the width of the wave produced by the Gaussian source, with $\eta \ll \pi$. The frequency of the incident wave is denoted by $\omega$. The function $X_{l0}(\theta)$ represents the spherical harmonics, while $C_{l0}$ denotes the expansion coefficients of $X_{l0}(\theta)$. These coefficients can be expressed as  
\begin{equation}
C_{l0} = (-1)^{l} \sqrt{\frac{l+1/2}{2\pi}} \, \mathrm{exp} \left[-\frac{(l+1/2)^{2} \eta ^{2} }{2} \right].
\end{equation}

Given the inherent symmetry of the spacetime, the scalar field $\tilde{\Psi}(\upsilon, y, \theta, \varphi)$ can be expressed as a mode expansion:  

\begin{equation}
\tilde{\Psi}(\upsilon, z, \theta, \varphi) = \sum_{l=0}^{\infty } \sum_{n=-l}^{l} e^{-i\omega \upsilon }C_{l0}U_{l}(z)
Y_{ln}(\theta ,\varphi ).  
\label{mode_expansion}
\end{equation}  

The corresponding response function $\left \langle \mathcal{Q} \right \rangle$ takes the form  

\begin{equation}
\left \langle \mathcal{Q} \right \rangle= \sum_{l=0}^{\infty } e^{-i\omega \upsilon }C_{l0}(\mathcal{Q})_{l} Y_{l0} (\theta ).
\label{response_function}
\end{equation}  

Using Eq. \eqref{mode_expansion}, the radial function $U_l(z)$ obeys the equation of motion  

\begin{eqnarray}
z^{2} N(z) {U}''_{l} + (z^{2} N'(z) - 2z N(z) + 2i\omega z^{2}) {U}'_{l} \nonumber\\  
+ (-2i\omega z - z^{2}l(l+1)) {U}_{l} = 0.
\label{radial_eq}
\end{eqnarray}  

Using the principles of AdS/CFT correspondence, near the AdS boundary, the solution for $U_{l}$ can be expanded as  \cite{holo13}

\begin{equation}
\lim_{z \to 0} U_{l} =1 - i\omega z + \frac{z^{2}(-l(1+l)) }{2} + \left \langle \mathcal{O} \right \rangle_{l} z^{3} + Q(z^{4}).
\end{equation}  

Clearly, the function $U_{l}$ is subject to two boundary conditions: one at the AdS boundary and the other at the event horizon, $z = z_h$.  At $z = 0$, which corresponds to the AdS boundary, the wave source $J_{\mathcal{Q} }$ represents the asymptotic form of the scalar field at infinity. From Eq. (\ref{radial_eq}), it follows that $U_{l}(0) = 1$. The function $U_{l}$ also satisfies a boundary condition at the event horizon $z = z_h$, given by  

\begin{equation}
(z_{h}^{2} N' + 2i\omega z_{h}^{2}) {U}'_{l} - [(2i\omega z_{h} + z_{h}^{2} l(l+1))]{U}_{l} = 0.
\end{equation}  
By applying these two boundary conditions, we numerically solve for $U_{l}$. Subsequently, the value of $\left \langle \mathcal{Q} \right \rangle_{l}$ is extracted using the pseudo-spectral method \cite{prl}. Finally, employing Eq. (\ref{response_function}), we obtain the total response function.\\

Figure \ref{2} illustrates the behavior of the response function $\left \langle \mathcal{Q} \right \rangle$ under different parameter variations. Fig.  \ref{2a} shows the influence of the  parameter $\alpha$ on the response function while keeping $a=1$, $z_{h}=1$, and $\omega=90$ fixed. It is evident that as $\alpha$ increases, the amplitude of the response function also increases, with the peaks becoming more pronounced.  Fig.  \ref{2b} presents the effect of varying the wave source frequency $\omega$ while maintaining $a=1$, $z_{h}=1$, and $\beta=0.01$. The results indicate that as $\omega$ increases, the periodicity of the response function decreases.Additionally, the amplitude of the peaks decreases with increasing $\omega$, suggesting a damping effect at higher frequencies. Finally, Fig.  \ref{2c} demonstrates how temperature $T$ influences the response function for fixed values of $a=1$, $\omega=90$, and $\beta=0.01$.  By selecting  values of $z_h$ as $1,2$ and $3$, we compute the corresponding temperature from eq.\eqref{temp} and analyze its effect on the response function. As depicted in Fig.  \ref{2c}, the amplitude of the response function decreases as temperature increases. 

\begin{figure}[h!]	
		\centering
		\begin{subfigure}{0.4\textwidth}
			\includegraphics[width=\linewidth]{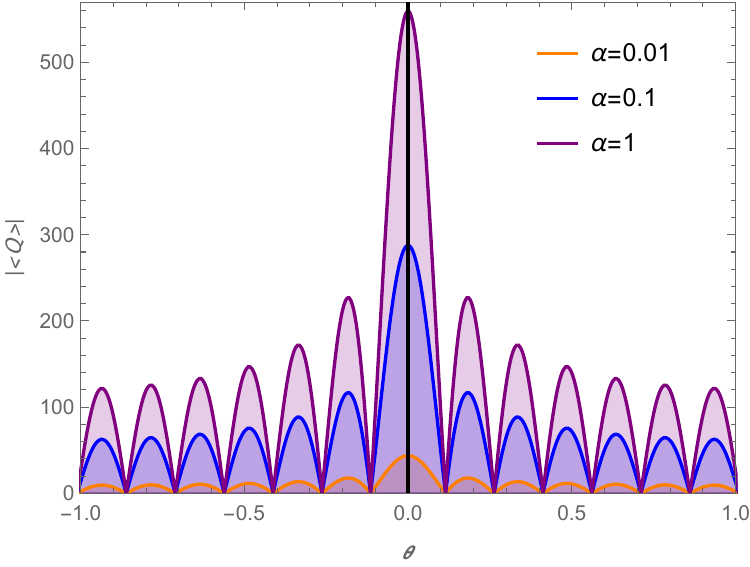}
			\caption{}
			\label{2a}
		\end{subfigure}
		\begin{subfigure}{0.4\textwidth}
			\includegraphics[width=\linewidth]{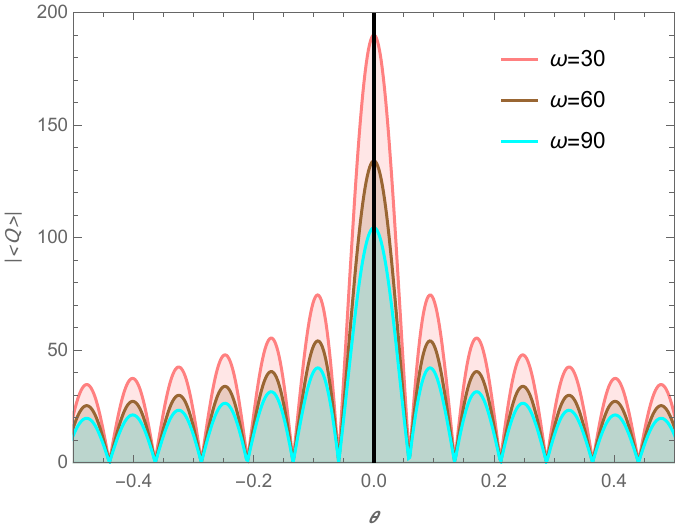}
			\caption{}
			\label{2b}
		\end{subfigure}
		\begin{subfigure}{0.4\textwidth}
			\includegraphics[width=\linewidth]{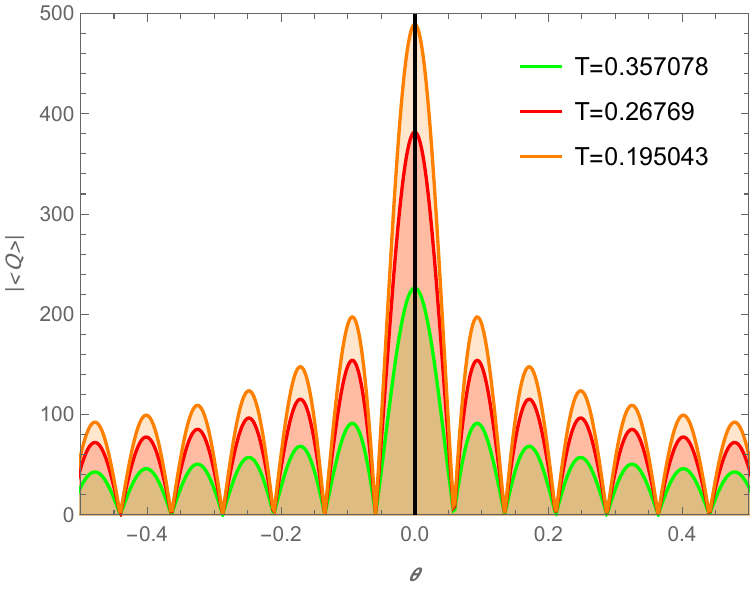}
			\caption{}
			\label{2c}
		\end{subfigure}
		\caption{Fig.(a) shows the effect of different  $\alpha$  on the  response function, where $\beta=1$, $z_{h}=1$, $\omega =90$.Fig. (b)illustrates the effect of different $\omega$ on the response function, where $\beta=1$, $z_{h}=1$, $\alpha=0.01$ and Fig.(c) represents the effect of different $T$  on the  response function, where $\beta=1$, $\omega=90$, $\alpha=0.01$.}
		\label{2}
	\end{figure}

 \section{The formation of holographic ring}

After deriving the total response function, we project it onto a spherical screen. As depicted in Fig. 2(b), the lens, with its focal points at $z = \pm f$, is considered to be infinitely thin, where the focal length satisfies $f \gg d$. When the response function is replicated as a plane wave around the observation point, we obtain $\Psi_p = \langle Q \rangle$. However, this function alone does not generate a holographic image. To achieve this, an optical setup incorporating a convex lens is necessary. The convex lens transforms an incident plane wave into a spherical wave, with the observation angle at the AdS boundary given by $\theta_{obs}$. A coordinate transformation from $(\theta, \varphi)$ to $(\theta', \varphi')$ is performed, satisfying the relation \cite{holo13}
\begin{equation}
\cos\varphi ' + i\cos\theta ' = e^{i\theta_{obs}} (\sin\theta \cos\varphi + i\cos\theta),
\end{equation}
where $\theta' = 0, \varphi' = 0$ corresponds to the center of observation. A Cartesian coordinate system $(x,y,z)$ is introduced, with the boundary observer's position given by $(x,y) = (\theta' \cos\varphi', \theta' \sin\varphi')$. The convex lens is placed on a two-dimensional plane $(x,y)$, where $f$ and $d$ denote its focal length and radius, respectively. The spherical screen coordinates are defined as $(x,y,z) = (x_{SC}, y_{SC}, z_{SC})$, satisfying the equation $f=\sqrt{x^2_{SC} + y^2_{SC} + z^2_{SC} }$. The relation between the incident wave $\Psi(\tilde{x})$ before passing through the convex lens and the outgoing wave $\Psi_T(\tilde{x})$ after transmission through the lens is given by
\begin{equation}
\Psi_T(\tilde{x}) = e^{-i\hat{\omega} \frac{|\tilde{x}|}{f} } \Psi(\tilde{x}).
\end{equation}
using the Fresnel approximation $f>>|x|$ the wave function on the screen can be represented as\cite{prl,holo13} \begin{eqnarray}
\Psi _{SC} (\tilde{x}_{SC}  )&=&\int d^{2} x\Psi (\tilde{x} )\sigma (\tilde{x})e^{-i\frac{\omega}{2f},
  \tilde{x}\cdot \tilde{x}_{SC} },\label{main}
\end{eqnarray} where $\sigma (\tilde{x})$ is defined as the window function, which is given by
 \begin{equation}
\sigma (\tilde{x}): = 
\begin{cases}
  1,~~~~0\le\tilde{x} \le d;  \\
  0,~~~~~~~~~~~\tilde{x} \ge  d.
\end{cases} 
\end{equation}

Furthermore, from eq.~(\ref{main}), the observed wave on the screen is related to the incident wave via the Fourier transform. In this work, the response function is treated as the incident wave $\Psi(\tilde{x})$, allowing the visualization of holographic images on the screen. The influence of different parameters associated with the black hole solution, on these holographic images is examined, with fixed source width $\eta = 0.02$ and convex lens radius $d = 1.2$. Consequently, dual black hole images can be captured on the screen, and by selecting appropriate black hole parameters, holographic Einstein rings can be obtained and analyzed.\\

Figure \ref{3} illustrates how the observational response appears on the screen for different values of the parameter $\alpha$ at various observation angles $\theta_{\text{obs}}$ while keeping $z_h=1$ and $\omega=40$ fixed. The intensity distribution is represented using a red-black and white colormap, where brighter regions indicate higher intensity, while darker regions correspond to lower intensity. The response is examined for three values of $\alpha$: 0.1 (top row), 0.05 (middle row), and 0.01 (bottom row), with the observation angle $\theta_{\text{obs}}$ varying from 0 to $\pi/2$ across the columns.  For $\theta_{\text{obs}} = 0$ (first column), the observed pattern takes the form of a nearly perfect circular ring, resembling a holographic Einstein ring. The radius of this ring changes  with change in $\alpha$ values. As the observation angle $\theta_{\text{obs}}$ increases (moving toward the right), the ring gradually deforms asymmetrically, transitioning from a full ring to a crescent-like shape. By the time $\theta_{\text{obs}} = \pi/2$ (last column), the ring is no longer continuous but instead appears asa  bright circular  spot.  When comparing different values of $\alpha$, we notice that a larger value i.e $\alpha-0.1$  results in a more distinct and well-defined Einstein ring, whereas as $\alpha$ value decreases,  it tends to produces a fainter but still visible structure. 
Overall, the figure highlights how both  $\alpha$ and the observation angle $\theta_{\text{obs}}$ significantly influence the appearance of holographic images. At lower angles, the Einstein ring remains largely intact, while at higher angles, it distorts into arcs and eventually reduces to isolated point like structure. 

\begin{figure*}[htbp]
\centering

\begin{subfigure}[b]{0.24\textwidth}
  \centering
\includegraphics[width=\textwidth]{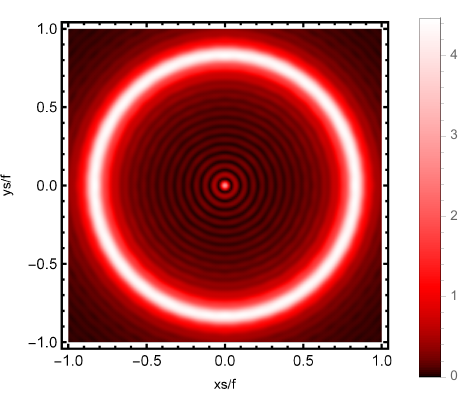}
  \caption{$\alpha=0.1$,$\theta _{obs}=0$}
\end{subfigure}
\hfill
\begin{subfigure}[b]{0.24\textwidth}
  \centering
  \includegraphics[width=\textwidth]{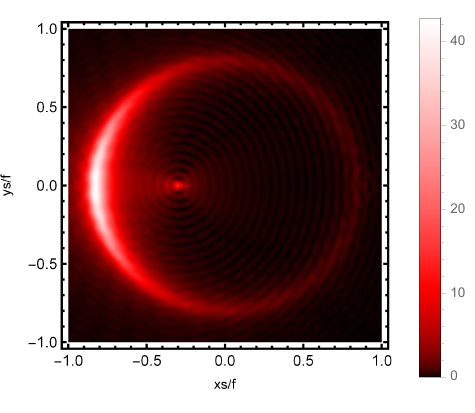}
  \caption{$\alpha=0.1$,$\theta _{obs}=\pi/6$}
\end{subfigure}
\hfill
\begin{subfigure}[b]{0.24\textwidth}
  \centering
  \includegraphics[width=\textwidth]{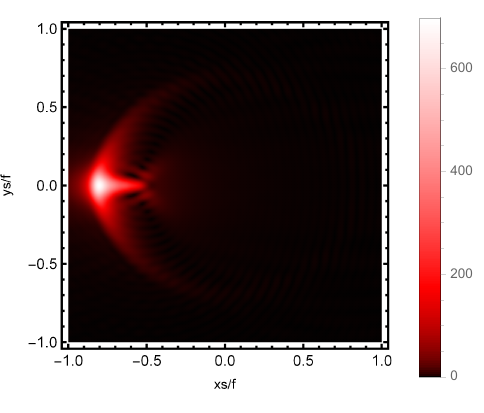}
  \caption{$\alpha=0.1$,$\theta _{obs}=\pi/3$}
\end{subfigure}
\hfill
\begin{subfigure}[b]{0.24\textwidth}
  \centering
  \includegraphics[width=\textwidth]{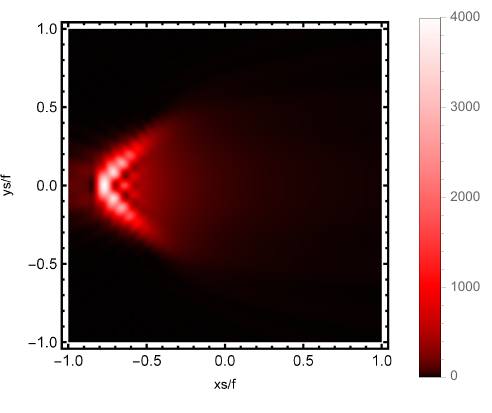}
  \caption{$\alpha=0.1$,$\theta _{obs}=\pi/2$}
\end{subfigure}
\hfill
\begin{subfigure}[b]{0.24\textwidth}
  \centering
  \includegraphics[width=\textwidth]{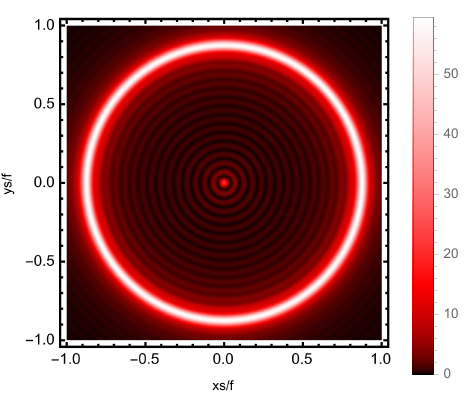} \caption{$\alpha=0.08$,$\theta _{obs}=0$}
\end{subfigure}
\hfill
\begin{subfigure}[b]{0.24\textwidth}
  \centering
  \includegraphics[width=\textwidth]{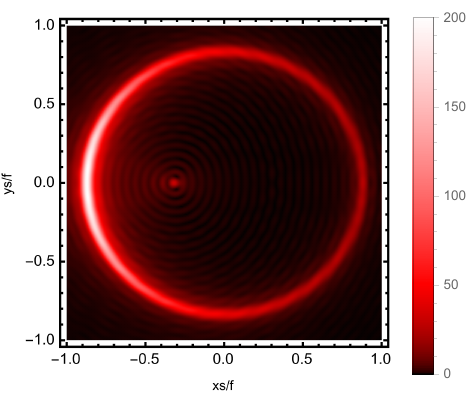}
  \caption{$\alpha=0.08$,$\theta _{obs}=\pi/6$}
\end{subfigure}
\hfill
\begin{subfigure}[b]{0.24\textwidth}
  \centering
  \includegraphics[width=\textwidth]{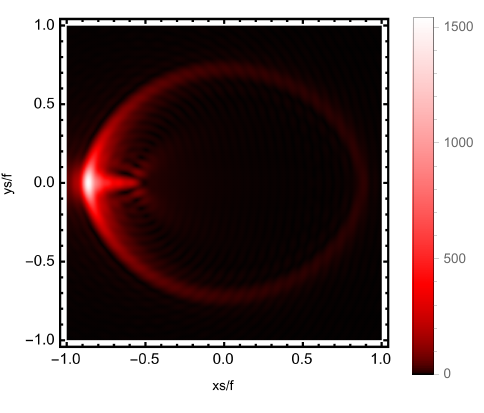}
  \caption{$\alpha=0.08$,$\theta _{obs}=\pi/3$}
\end{subfigure}
\hfill
\begin{subfigure}[b]{0.24\textwidth}
  \centering
  \includegraphics[width=\textwidth]{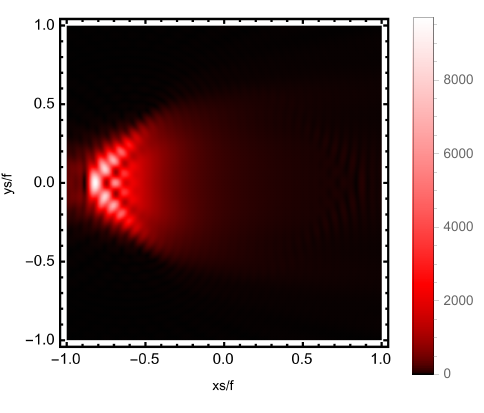}
  \caption{$\alpha=0.08$,$\theta _{obs}=\pi/2$}
\end{subfigure}
\hfill
\begin{subfigure}[b]{0.24\textwidth}
  \centering
\includegraphics[width=\textwidth]{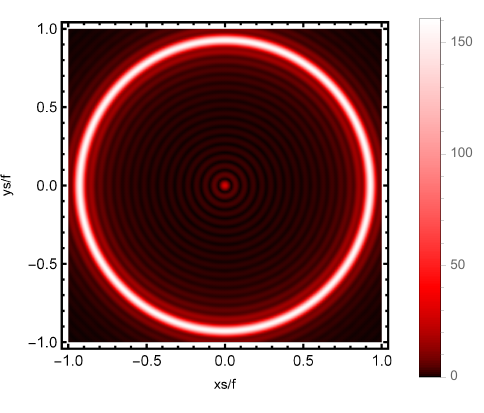} \caption{$\alpha=0.05$,$\theta _{obs}=0$}
\end{subfigure}
\hfill
\begin{subfigure}[b]{0.24\textwidth}
  \centering
  \includegraphics[width=\textwidth]{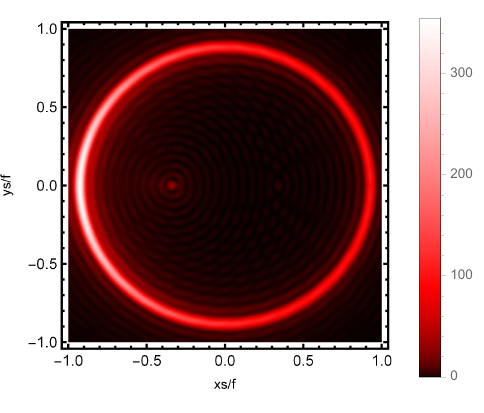}
  \caption{$\alpha=0.05$,$\theta _{obs}=\pi/6$}
\end{subfigure}
\hfill
\begin{subfigure}[b]{0.24\textwidth}
  \centering
  \includegraphics[width=\textwidth]{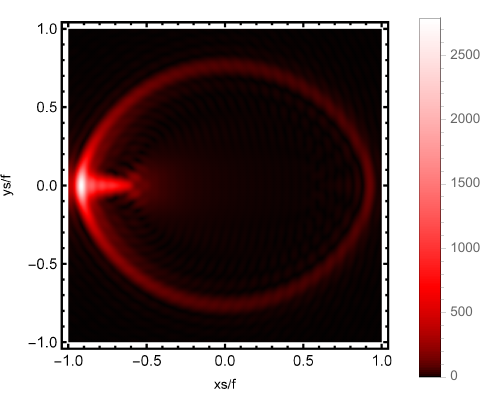}
  \caption{$\alpha=0.05$,$\theta _{obs}=\pi/3$}
\end{subfigure}
\hfill
\begin{subfigure}[b]{0.24\textwidth}
  \centering
  \includegraphics[width=\textwidth]{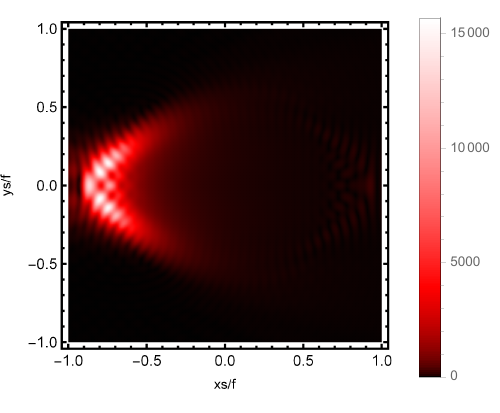}
  \caption{$\alpha=0.05$,$\theta _{obs}=\pi/2$}
\end{subfigure}
\caption{Observational appearance of the response on the screen for different $\alpha$ at various observation angles, where $\beta=1,z_{h}=1$, $\omega =40$.}
\label{3}%
\end{figure*}
To further analyze the influence of the parameter $\alpha$ on the holographic image, Figure \ref{3} presents both the intensity distribution (top row) and the corresponding brightness profiles (bottom row). The top row illustrates the observed Einstein ring under different values of $\alpha$ where we have kept $z_h=1,\omega=40$ and $\theta_{obs}=0$ constant. The bottom row shows the brightness variation along the horizontal axis of the image. The brightness profiles reveal that the intensity peaks correspond to the locations of the Einstein ring. The separation between the two dominant peaks in each brightness plot represents the effective diameter of the ring. As $\alpha$ increases, the radius of the Einstein ring exhibits a slight expansion, indicating that the $\alpha$ parameter influences the size of the observed structure. For instance, at $\alpha = 0.1$, the ring appears more pronounced, whereas at $\alpha = 0.05$, it becomes relatively thinner. This suggests that higher values of $\alpha$ lead to a more expanded and well-defined ring.  Furthermore, the peak intensity decreases for larger values of $\alpha$.So it can be concluded  that $\alpha$ not only governs the size of the Einstein ring but also affects its brightness distribution. A higher value of $\alpha$ parameter produces a larger yet fainter ring, while a lower $\alpha$ results in a more compact but sharper structure. \\

The effect of frequency $\omega$ on the holographic image is illustrated in Fig.~\ref{4} where we have kept $\alpha=0.1,z_h=1$ and $\theta_{obs}=0$ constant while varying the $\omega$. The holographic images reveal a series of concentric patterns. For $\omega=40$ and $\omega=70$, the central region of the ring appears dark, but in reality, a faint bright spot exists at the center. Its visibility, however, depends on parameter choices, making it nearly imperceptible in some cases. Under specific conditions, this central brightness becomes more pronounced, resembling the Poisson–Arago spot.  At lower values of $\omega$,  the image primarily consists of a central bright spot surrounded by faint concentric rings,  However, as $\omega$  increases, these rings become more pronounced, and a well-defined bright circular ring emerges around the center. 
The corresponding brightness profiles further illustrate this behavior. For small $\omega$ , the brightness distribution along the \(x_{\text{slf}}\) axis is dominated by a single central peak, with relatively weak oscillations on either side. As \(\omega\) increases, the distribution becomes more intricate, with multiple sharp peaks appearing at well-defined locations.  With increasing $\omega$, the intensity peak is observed to be decreasing. Additionally, the radius of the bright ring expands with increasing $\omega$.  


\begin{figure*}[t!]
\centering
\begin{subfigure}[b]{0.3\textwidth}
  \centering
  \includegraphics[width=\textwidth]{A5.pdf}
  \caption{$\alpha=0.1$}
\end{subfigure}
\hfill
\begin{subfigure}[b]{0.3\textwidth}
  \centering
  \includegraphics[width=\textwidth]{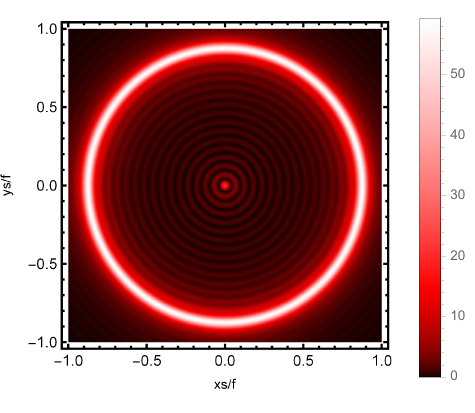}
  \caption{$\alpha=0.08$}
\end{subfigure}
\hfill
\begin{subfigure}[b]{0.3\textwidth}
  \centering
  \includegraphics[width=\textwidth]{A9.pdf}
  \caption{$\alpha=0.05$}
\end{subfigure}

\vspace{0.5cm} 

\begin{subfigure}[b]{0.3\textwidth}
  \centering
  \includegraphics[width=\textwidth]{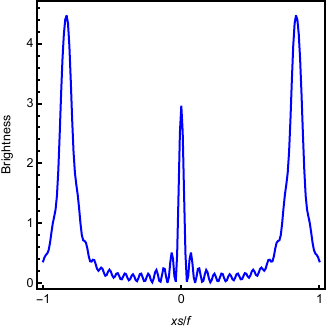}
  \caption{$\alpha=0.1$}
\end{subfigure}
\hfill
\begin{subfigure}[b]{0.3\textwidth}
  \centering
  \includegraphics[width=\textwidth]{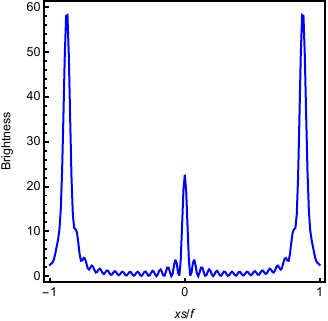}
  \caption{$\alpha=0.08$}
\end{subfigure}
\hfill 
\begin{subfigure}[b]{0.3\textwidth}
  \centering
  \includegraphics[width=\textwidth]{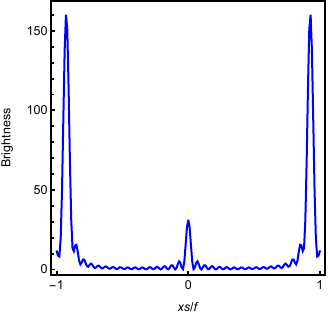}
  \caption{$\alpha=0.05$}
\end{subfigure}

\caption{Effect of $\alpha$ on the holographic image where $\beta=1$, $\omega=50$, $z_h=1$, and $\theta_{\text{obs}}=0$.}
\label{3}
\end{figure*}

Next, we analyze the effect of temperature on the holography image.   Fig.\ref{5},  illustrates the holographic images and their corresponding three-dimensional brightness profiles for different $z_h$ values. As the parameter $z_h$ increases, subtle variations in the size and brightness distribution of these rings become apparent. In the first two row (sub-figure (a)-(h))of the images in Fig.\ref{5}, demonstrate the effect of temperature, while keeping the observer's inclination fixed at \( \theta_{\text{obs}} = 0 \). From the temperature expression, it is evident that the temperature is inversely proportional to \( z_h \), meaning that larger values of \( z_h \) correspond to lower temperatures. The top row presents the intensity maps of the holographic images, which display concentric ring structures.  As \( z_h \) increases, the peak brightness of the rings becomes more pronounced. In other words, the peak intensity of the holographic image decreases with increasing temperature.

The second row (sub-figure (e)-(h)) shows the three-dimensional brightness profiles corresponding to each holographic image, where the three axes represent key observational properties. The horizontal ($ x_{s/f}$) and vertical ($ y_{s/f}$) axes define the observer’s image plane, while the vertical axis represents the brightness or intensity of the radiation received by the observer. These plots provide a quantitative representation of the brightness variations, with sharp peaks signifying regions of maximum intensity. As $z_h$ increases, the peak brightness also increases, suggesting that lower temperatures enhance the contrast and sharpness of the holographic image. \\
The same analysis has been repeated for $\theta_{obs}=\pi/4$ in third and fourth row of Fig.\ref{5} (sub-figure (i)-(p)) which illustrate the holographic images and their corresponding three-dimensional brightness profiles for different values of $z_h$ . Here also as \( z_h \) increases, the brightness distribution evolves, and the peak intensity of the rings becomes more pronounced. This suggests that lower temperatures enhance the contrast and sharpness of the observed structure.

A similar analysis is presented in fifth and sixth row(subfigures (q)-(x)) where $\theta_{obs}$. is set equal to $\pi/2$ The qualitative trends remain similar, with concentric ring structures in the intensity maps (fifth row) and corresponding three-dimensional brightness profiles (sixth row). The contrast and peak intensity follow the same increasing trend with \( z_h \), highlighting the role of both temperature and inclination angle in shaping the observed features of the holographic system.

\begin{figure*}[h!]
\centering
\begin{subfigure}[b]{0.31\textwidth}
  \includegraphics[width=\textwidth]{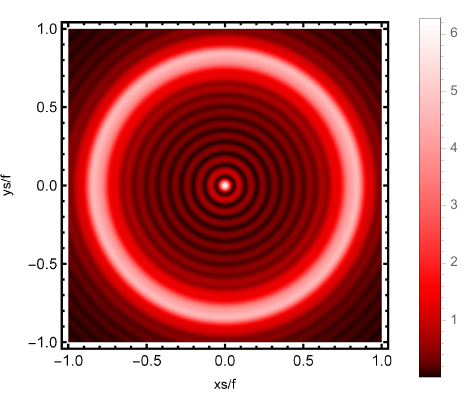}
  \caption{$\omega=30$}
\end{subfigure}
\hfill
\begin{subfigure}[b]{0.31\textwidth}
  \includegraphics[width=\textwidth]{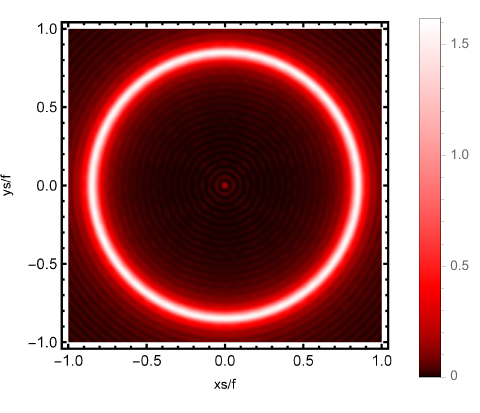}
  \caption{$\omega=50$}
\end{subfigure}
\hfill
\begin{subfigure}[b]{0.31\textwidth}
  \includegraphics[width=\textwidth]{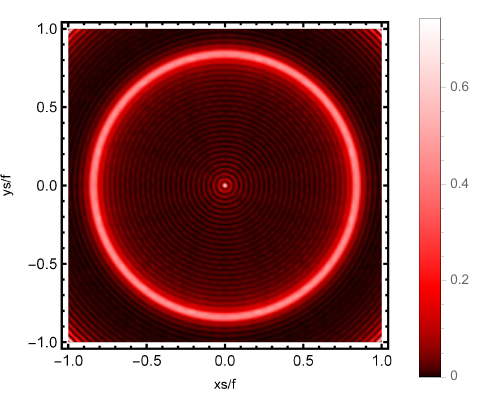}
  \caption{$\omega=70$}
\end{subfigure}

\vspace{0.5cm} 

\begin{subfigure}[b]{0.3\textwidth}
  \includegraphics[width=\textwidth]{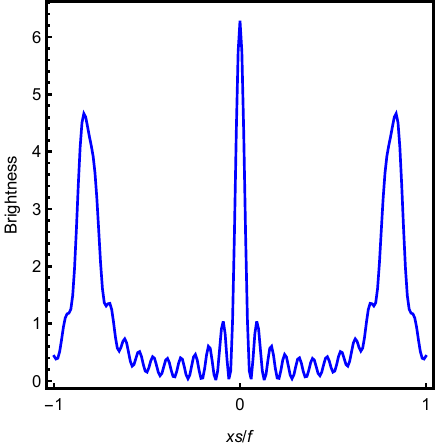}
  \caption{$\omega=30$}
\end{subfigure}
\hfill
\begin{subfigure}[b]{0.3\textwidth}
  \includegraphics[width=\textwidth]{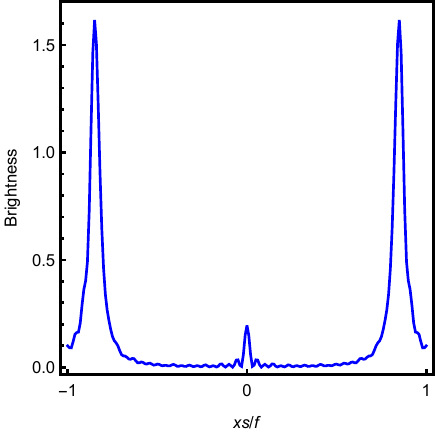}
  \caption{$\omega=50$}
\end{subfigure}
\hfill
\begin{subfigure}[b]{0.3\textwidth}
 \includegraphics[width=\textwidth]{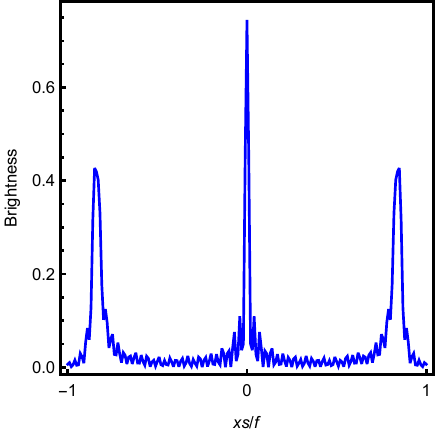}
  \caption{$\omega=70$}
\end{subfigure}

\caption{Effect of $\omega$ on the holographic image where $\alpha=0.1$, $\beta=1$, $z_h=1$, and $\theta_{\text{obs}}=0$.}
\label{4}
\end{figure*}

\begin{figure*}[htbp]
\centering
\begin{subfigure}[b]{0.24\textwidth}
  \centering
\includegraphics[width=\textwidth]{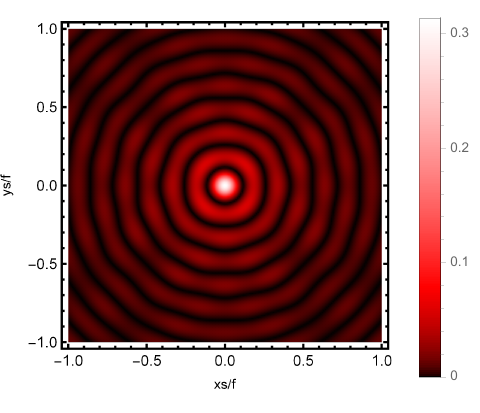}
  \caption{$z_h=0.5,\theta_{obs}=0$}
\end{subfigure}
\hfill
\begin{subfigure}[b]{0.24\textwidth}
  \centering
  \includegraphics[width=\textwidth]{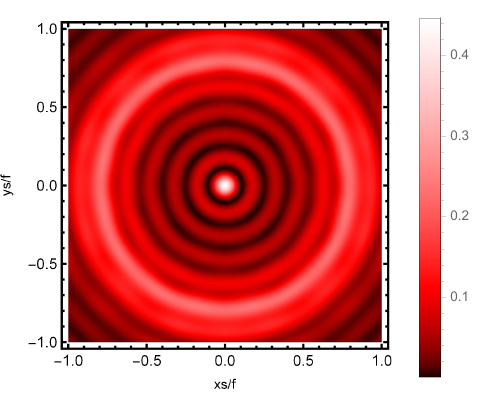}
  \caption{$z_h=1,\theta_{obs}=0$}
\end{subfigure}
\hfill
\begin{subfigure}[b]{0.24\textwidth}
  \centering
  \includegraphics[width=\textwidth]{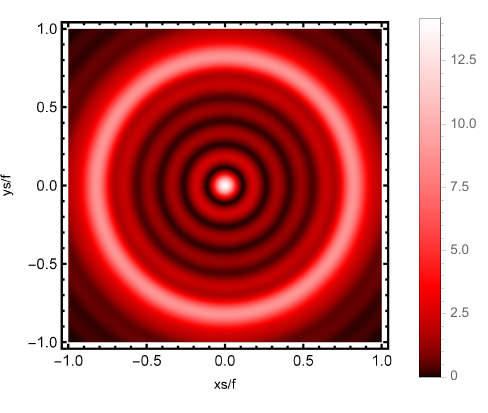}
  \caption{$z_h=1.5,\theta_{obs}=0$}
\end{subfigure}
\hfill
\begin{subfigure}[b]{0.24\textwidth}
  \centering
  \includegraphics[width=\textwidth]{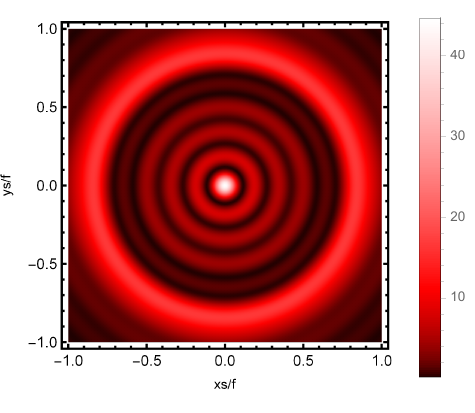}
  \caption{$z_h=2,\theta_{obs}=0$}
\end{subfigure}
\hfill
\begin{subfigure}[b]{0.24\textwidth}
  \centering
\includegraphics[width=\textwidth]{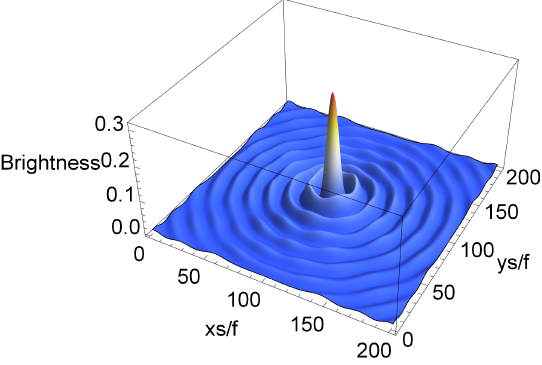}
  \caption{$z_h=0.5,\theta_{obs}=0$}
\end{subfigure}
\hfill
\begin{subfigure}[b]{0.24\textwidth}
  \centering
  \includegraphics[width=\textwidth]{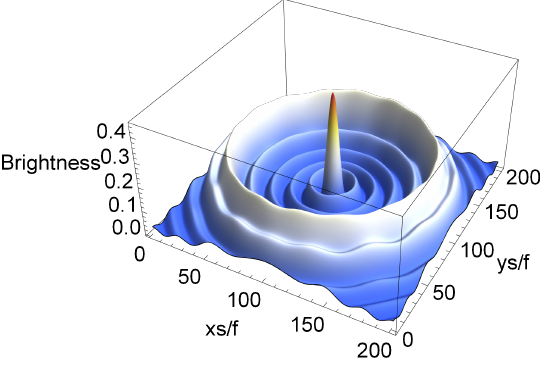}
  \caption{$z_h=1,\theta_{obs}=0$}
\end{subfigure}
\hfill
\begin{subfigure}[b]{0.24\textwidth}
  \centering
  \includegraphics[width=\textwidth]{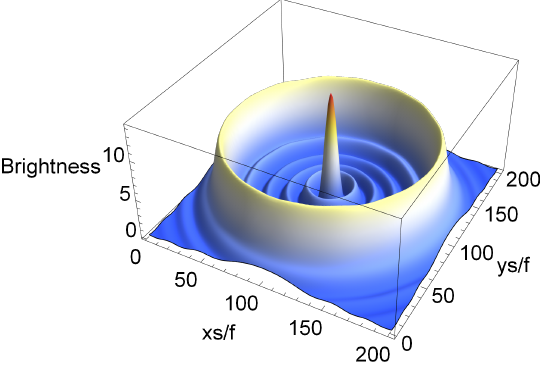}
  \caption{$z_h=1.5,\theta_{obs}=0$}
\end{subfigure}
\hfill
\begin{subfigure}[b]{0.24\textwidth}
  \centering
  \includegraphics[width=\textwidth]{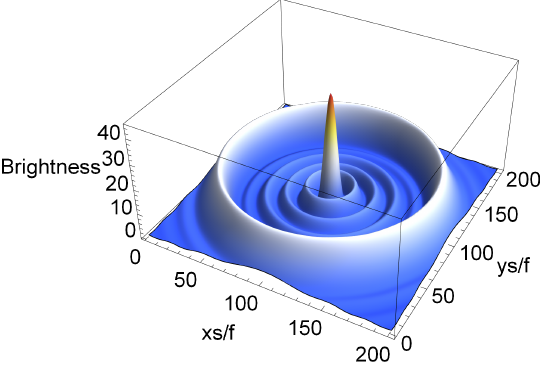}
  \caption{$z_h=2,\theta_{obs}=0$}
\end{subfigure}
\hfill
\begin{subfigure}[b]{0.24\textwidth}
\includegraphics[width=\textwidth]{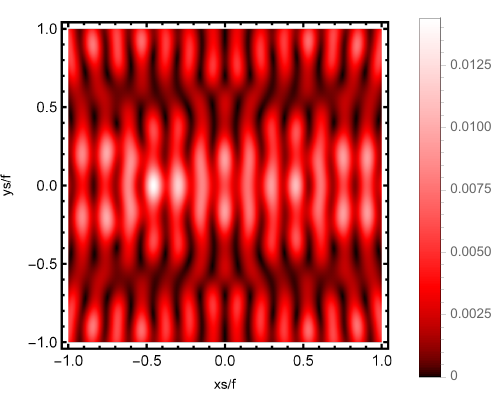}
  \caption{$z_h=0.5,\theta_{obs}=\frac{\pi}{4}$}
\end{subfigure}
\hfill
\begin{subfigure}[b]{0.24\textwidth}
  \centering
  \includegraphics[width=\textwidth]{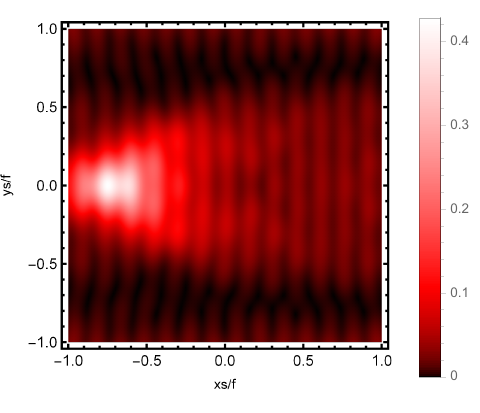}
  \caption{$z_h=1,\theta_{obs}=\frac{\pi}{4}$}
\end{subfigure}
\hfill
\begin{subfigure}[b]{0.24\textwidth}
  \centering
  \includegraphics[width=\textwidth]{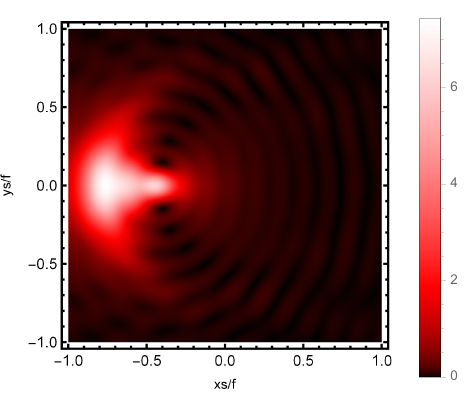}
  \caption{$z_h=1.5,\theta_{obs}=\frac{\pi}{4}$}
\end{subfigure}
\hfill
\begin{subfigure}[b]{0.24\textwidth}
  \centering
  \includegraphics[width=\textwidth]{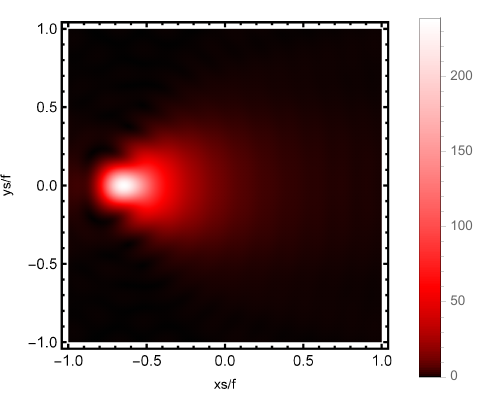}
  \caption{$z_h=2,\theta_{obs}=\frac{\pi}{4}$}
\end{subfigure}
\hfill
\begin{subfigure}[b]{0.24\textwidth}
  \centering
\includegraphics[width=\textwidth]{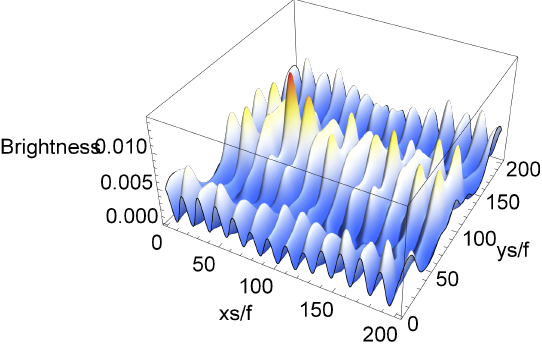}
  \caption{$z_h=0.5,\theta_{obs}=\frac{\pi}{4}$}
\end{subfigure}
\hfill
\begin{subfigure}[b]{0.24\textwidth}
  \centering
  \includegraphics[width=\textwidth]{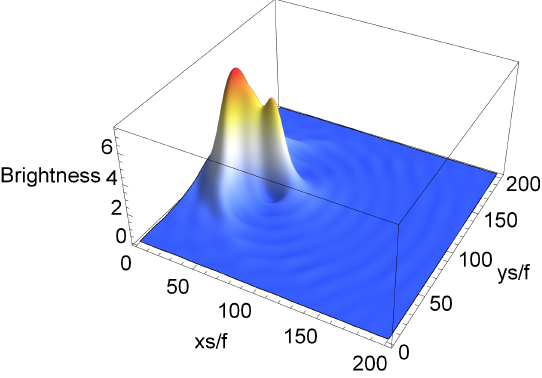}
  \caption{$z_h=1,\theta_{obs}=\frac{\pi}{4}$}
\end{subfigure}
\hfill
\begin{subfigure}[b]{0.24\textwidth}
  \centering
  \includegraphics[width=\textwidth]{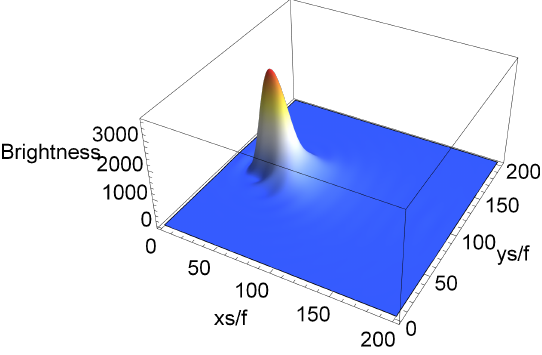}
  \caption{$z_h=1.5,\theta_{obs}=\frac{\pi}{4}$}
\end{subfigure}
\hfill
\begin{subfigure}[b]{0.24\textwidth}
  \centering
  \includegraphics[width=\textwidth]{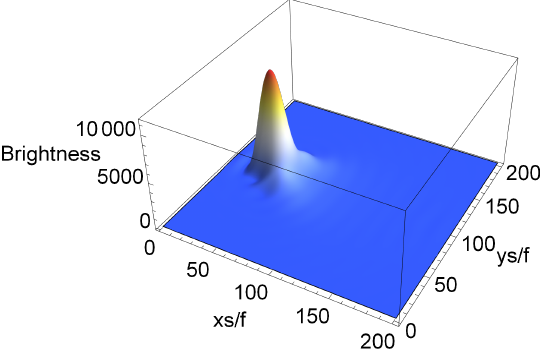}
  \caption{$z_h=2,\theta_{obs}=\frac{\pi}{4}$}
\end{subfigure}
\hfill
\begin{subfigure}[b]{0.24\textwidth}
\includegraphics[width=\textwidth]{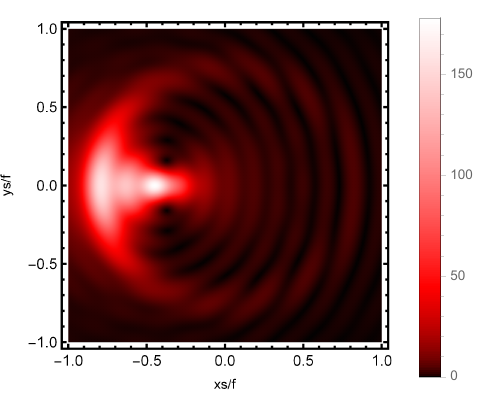}
  \caption{$z_h=0.5,\theta_{obs}=\frac{\pi}{2}$}
\end{subfigure}
\hfill
\begin{subfigure}[b]{0.24\textwidth}
  \centering
  \includegraphics[width=\textwidth]{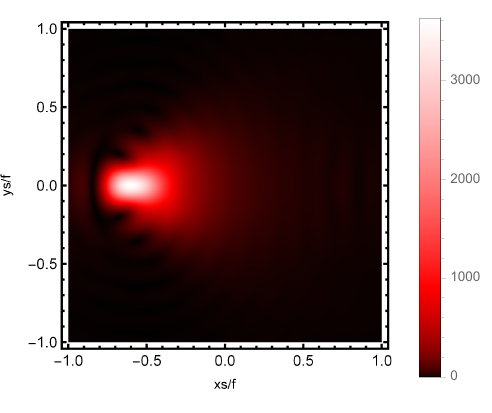}
  \caption{$z_h=1, \theta_{obs}=\frac{\pi}{2}$}
\end{subfigure}
\hfill
\begin{subfigure}[b]{0.24\textwidth}
  \centering
  \includegraphics[width=\textwidth]{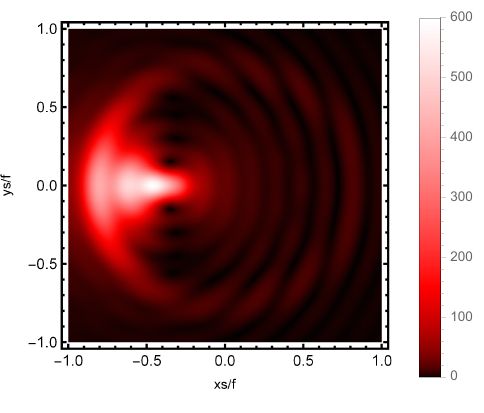}
  \caption{$z_h=1.5,\theta_{obs}=\frac{\pi}{2}$}
\end{subfigure}
\hfill
\begin{subfigure}[b]{0.24\textwidth}
  \centering
  \includegraphics[width=\textwidth]{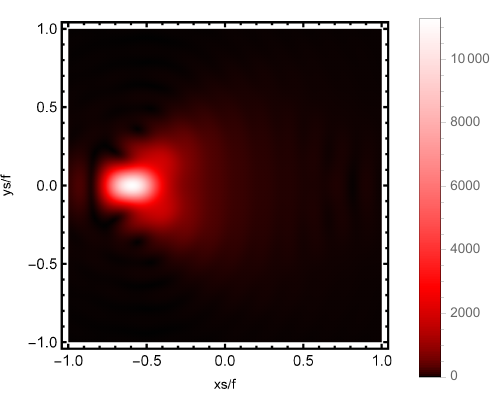}
  \caption{$z_h=2,\theta_{obs}=\frac{\pi}{2}$}
\end{subfigure}
\hfill
\begin{subfigure}[b]{0.24\textwidth}
  \centering
\includegraphics[width=\textwidth]{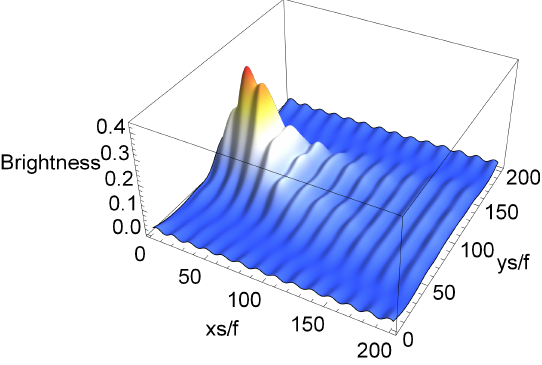}
  \caption{$z_h=0.5,\theta_{obs}=\frac{\pi}{2}$}
\end{subfigure}
\hfill
\begin{subfigure}[b]{0.24\textwidth}
  \centering
  \includegraphics[width=\textwidth]{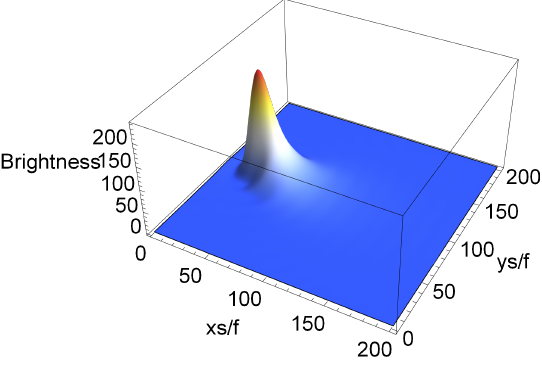}
  \caption{$z_h=1,\theta_{obs}=\frac{\pi}{2}$}
\end{subfigure}
\hfill
\begin{subfigure}[b]{0.24\textwidth}
  \centering
  \includegraphics[width=\textwidth]{1.5zh90.pdf}
  \caption{$z_h=1.5,\theta_{obs}=\frac{\pi}{2}$}
\end{subfigure}
\hfill
\begin{subfigure}[b]{0.24\textwidth}
  \centering
  \includegraphics[width=\textwidth]{2zh90.pdf}
  \caption{$z_h=2,\theta_{obs}=\frac{\pi}{2}$}
\end{subfigure}
\hfill
\caption{Effect of $z_h$ on holographic image where $\alpha=0.1$,$\omega=40$}
\label{5}%
\end{figure*}
\clearpage

\section{Comparison between geometric and optical results}
In the context of black hole imagery, the luminous ring observed corresponds to the photon sphere of the black hole. To further explore the physical implications of the holographic Einstein ring, we will examine the relationship between this bright ring and the photon sphere from the standpoint of geometric optics. The Eikonal approximation is particularly useful in this analysis as it reformulates the Klein-Gordon equation into the Hamilton-Jacobi equation format and is given as \cite{holo13,holo11}-
\begin{equation}
g^{\mu \nu} (\partial _\mu \mathcal{I}-eA_t)(\partial_\nu-eA_t)-\mathbb{M}^2=0
\end{equation}
where the action is given as $\mathcal{I}$. From (17), the above equation is given as
\begin{equation}
-N(r)^{-1}(\partial_t \mathcal{I}-eA_t)^2+N(r)\partial_r\mathcal{I}^2+\frac{1}{r^2}\partial_\theta^2+\frac{1}{r^2\sin^2\theta}\partial^2_\varphi-\mathbb{M}=0
\end{equation}
The spherically symmetric structures described in Eq. (17) permit the placement of photon orbits exclusively on the equatorial plane, thus simplifying the model without loss of generality, whereby 
$\theta=\pi/2$ and $\dot{\theta}=0$ In this spacetime configuration, the presence of two Killing vectors,$\omega$ and $L$, facilitates the formulation of the action as follows \cite{holo11,holo13}.
\begin{equation}
\mathcal{I}(t,r,\varphi)=-\omega t+L\varphi+\int N(r)^{-1}\sqrt{\mathcal{R}}dr
\end{equation}
where $\mathcal{R}$ is given as \begin{equation}
\mathcal{R}=(\omega-eA_t)^2-N(r)(\frac{L^2}{r^2}-2)
\end{equation}
In which
\begin{equation}
\omega=-\frac{\partial \mathcal{I}}{\partial t}, \quad L=\frac{\partial \mathcal{I}}{\partial \varphi}, \quad \frac{\sqrt{\mathcal{R}}}{N(r)}=\frac{\partial \mathcal{I}}{\partial r}
\end{equation}
\begin{figure}[h]	
			\includegraphics[scale=0.3]{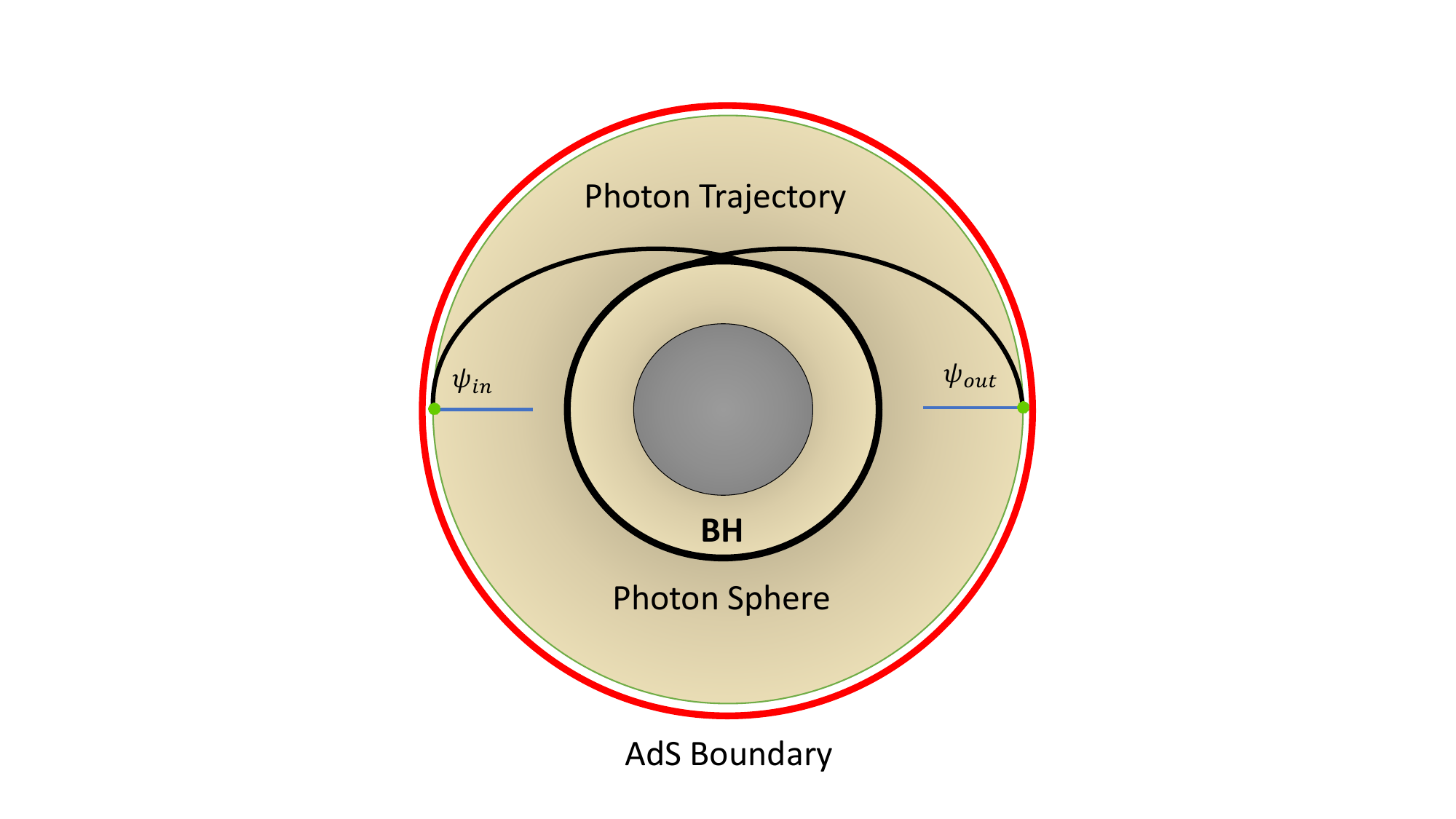}
			\caption{The primary contribution to the final response function arises from trajectories that closely approach the circular orbit.}
			\label{imag1}
		\end{figure}
In the context of the equation, $\omega$ and $L$ represent the conserved energy and angular momentum, respectively. The trajectory of photon propagation is dictated by these aforementioned conditions. Subsequently, the ingoing angle 
$\psi_{in}$ in , defined relative to the normal vector of the boundary 
$n^\gamma=\frac{\partial}{\partial r^\gamma}$ is characterized as follows \cite{prl,holo13}
\begin{equation}
\cos\psi_{in}=\frac{g_{ij}u^i n^j}{|u||n|}|_{r=\infty}=\sqrt{\frac{\dot{r}/N}{\dot{r}^2/N+L/r^2}}|_{r=\infty}
\end{equation}	
		\begin{figure}[h]
			\includegraphics[width=\linewidth]{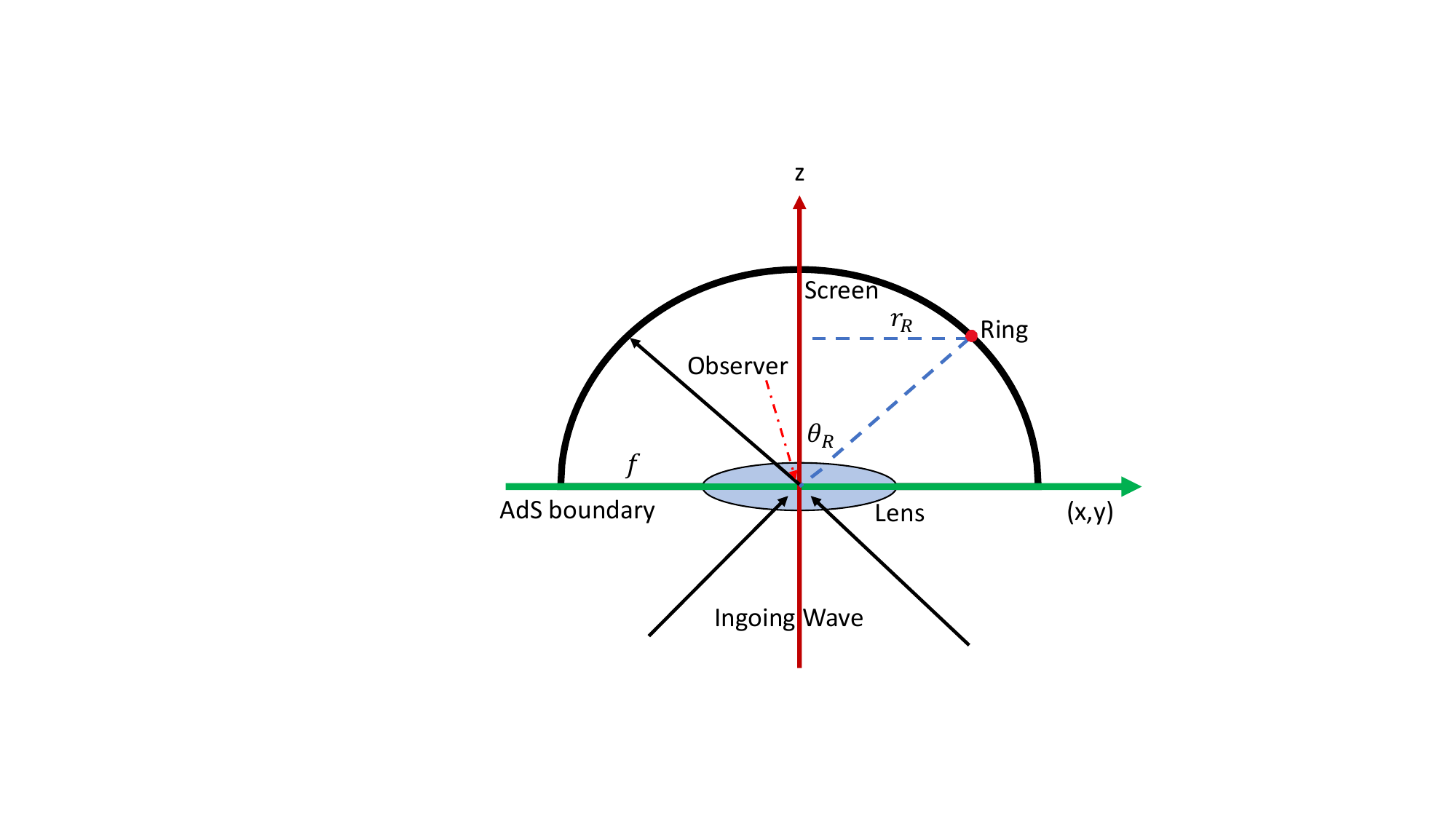}
			\caption{ Dependence of the Einstein ring radius $r_R$ on ring angle $\theta_R$.}
			\label{imag2}
		\end{figure}
 We can also write
 \begin{equation}
 \sin\psi^2_{in}=1-\cos\psi^2_{in}=\frac{L^2 N/r^2}{\dot{r}^2+L^2N/r^2}|_{r=\infty}=\frac{L^2}{\omega^2}
 \end{equation}
Therefore, the ingoing angle $\psi_{in}$ of the photon orbit from the boundary is consequently determined as
\begin{equation}
\label{eq:lo}
\sin\psi_{in}=\frac{L}{\omega}
\end{equation}
Photons near the photon sphere will orbit the black hole one or more times, leading to an increase in luminosity within this region. Photons found in proximity to the photon sphere are also subject to Eq. \eqref{eq:lo}, and the corresponding diagram is depicted in Figure \ref{imag1}. Here, the angular momentum off these photons is represented as $L_S$, and their light trajectories fulfil the following conditions\cite{prl,holo13}
\begin{equation}
\mathcal{R}=0, \quad \frac{d \mathcal{R}}{dr}=0
\end{equation}
From the standpoint of geometric optics, the angle $\psi_{in}$ denotes the angular displacement of the incident ray's image from the zenith as observed from the AdS boundary. In situations where the entry and exit points of photon trajectories at the AdS boundary are directly aligned with the black hole's center \cite{prl}, the radius of the ring image observed by an observer is determined by the incidence angle $\psi_{in}$ due to the axial symmetry. Consequently, the Einstein ring forms on the screen, as illustrated in Fig.\ref{imag2}, with the specified ring radius given as
\begin{equation}
\sin\theta_R=\frac{r_R}{f}
\end{equation}
We can get $\sin\theta_R=\sin\psi_{in}$ given in [33,34] for a large $\varsigma$ which gives
\begin{equation}
\frac{r_R}{f}=\frac{L_S}{\omega}
\end{equation}
\begin{figure}[h]	
		\centering
		\begin{subfigure}{0.4\textwidth}
			\includegraphics[width=\linewidth]{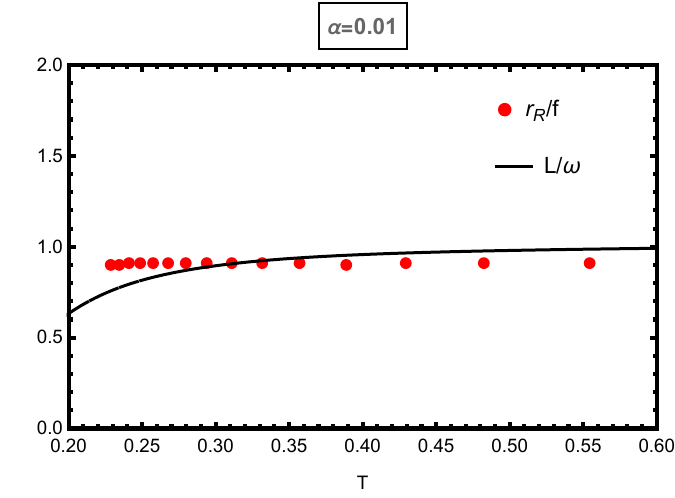}
			\caption{}
			\label{2a}
		\end{subfigure}
		\begin{subfigure}{0.4\textwidth}
			\includegraphics[width=\linewidth]{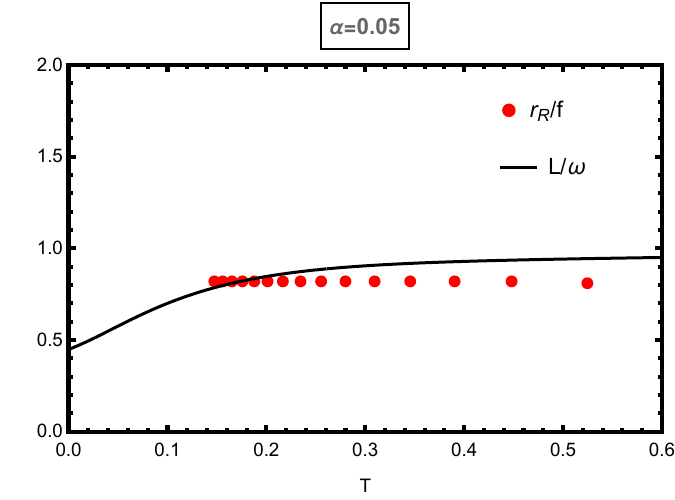}
			\caption{}
			\label{2b}
		\end{subfigure}
		\begin{subfigure}{0.4\textwidth}
			\includegraphics[width=\linewidth]{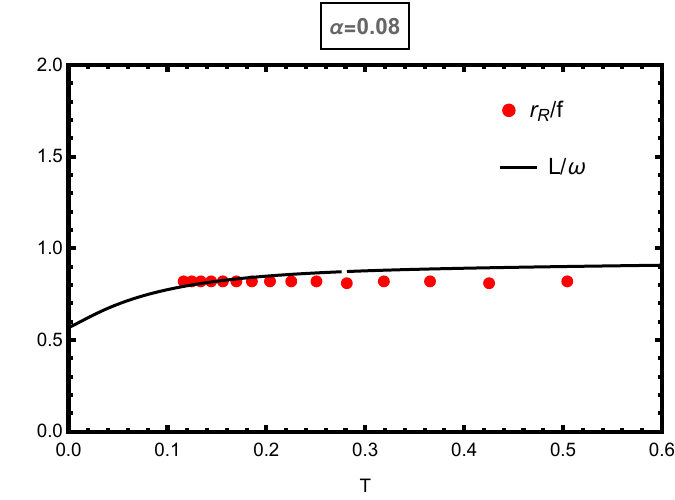}
			\caption{}
			\label{2c}
		\end{subfigure}
		\begin{subfigure}{0.4\textwidth}
			\includegraphics[width=\linewidth]{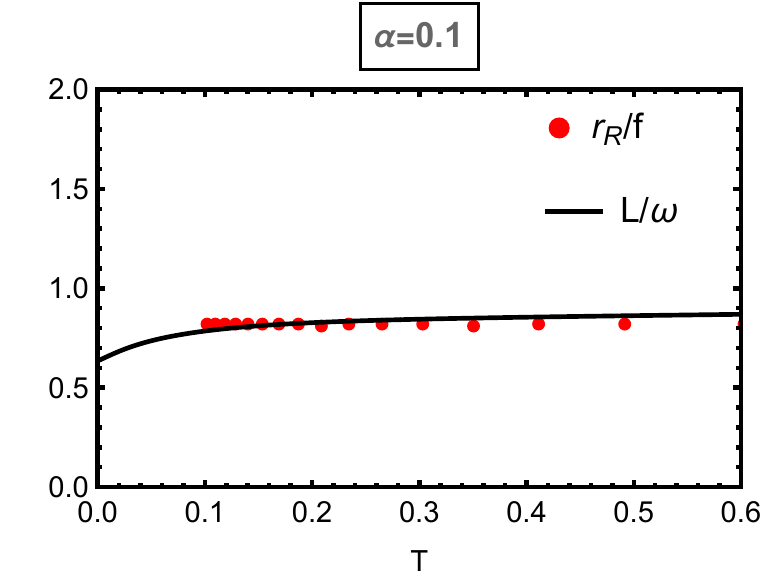}
			\caption{}
			\label{2c}
		\end{subfigure}
		\caption{The comparison of the Einstein ring radius between geometric optics and wave optics for different values of $\alpha$ with $\omega=50$. The discrete red points indicate the Einstein ring radius obtained from wave optics, whereas the black curve illustrates how the radius of the circular orbit varies with temperature, as determined from geometric optics.}
		\label{2}
	\end{figure}
The focal length $f$ of the lens determines the magnification level, thereby aligning the radius of the holographic ring with the size of the photon ring. Essentially, both the incident angle of the photon and the angle of the photon ring, as observed by an observer, refer to the same angular measurement and are fundamentally equivalent. This correlation will be substantiated through numerical analysis. In Figure \ref{2}, the radius of the Einstein ring is shown for varying values of the parameter $\alpha$ where $r_R/f$ is the ring radius expressed as a function of temperature and chemical potential. The alignment of the red data points closely follows the black curve, deviating by no more than three percent. The figures collectively demonstrate that the results obtained through wave optics via holography align with those derived from geometric optics, thus validating the holographic approach for studying black hole images. Further discussion of these comparative results reveals that the Einstein radius calculated from geodesic analysis, depicted by the blue curves, closely matches the Einstein radius constructed from the response function in wave optics. Small discrepancies may arise due to numerical precision and wave effects. This suggests that the predominant contribution to the brightest ring in such images originates from 'light rays" near the photon sphere, where they accumulate indefinitely. Any deviation of the Einstein radius from the geodesic predictions might be attributed to some wave effects. This consistency between the Einstein ring angle observed in holographic studies and the ingoing angle calculated through geometric optics demonstrates a strong concordance between these two methodologies.\\
\section{Conclusion}

In this work, we explored the holographic optical appearance of an AdS black hole with higher-derivative corrections in the presence of a string cloud, utilizing the AdS/CFT correspondence and wave optics techniques. By introducing a Gaussian wave source at the AdS boundary and analyzing the response function, we studied how the black hole’s optical features change under different physical parameters.\\

First, we analyzed the effect of the Gauss-Bonnet coupling constant $\alpha$ on the response function and found that as $\alpha$ increases, the amplitude of the response function also increases, with the peaks becoming more pronounced. Additionally, as the wave source frequency $\omega$ increases, the periodicity of the response function decreases, while the amplitude of the peaks diminishes, indicating a damping effect at higher frequencies. The impact of temperature $T$ was also examined, revealing that the amplitude of the response function decreases as temperature increases.\\

Next, we explored how the observational response appears on the screen for different values of $\alpha$ at various observation angles $\theta_{\text{obs}}$. For $\theta_{\text{obs}} = 0$, the observed image forms a nearly perfect circular ring, resembling a holographic Einstein ring. The radius of this ring varies with changes in $\alpha$. As the observation angle $\theta_{\text{obs}}$ increases, the ring progressively deforms asymmetrically, transitioning from a complete ring to a crescent-like shape. By the time $\theta_{\text{obs}} = \pi/2$, the ring is no longer continuous and instead appears as a bright circular spot. Comparing different values of $\alpha$, we find that larger values of $\alpha$ result in a more distinct and well-defined Einstein ring, while smaller $\alpha$ values produce a fainter but still visible structure. Furthermore, the peak intensity decreases as $\alpha$ increases. This suggests that $\alpha$ not only determines the size of the Einstein ring but also influences its brightness distribution. A higher $\alpha$ produces a larger yet fainter ring, whereas a lower $\alpha$ results in a more compact but sharper structure.  We also investigated the effect of frequency $\omega$ on the holographic image. As $\omega$ increases, the rings become more distinct, and a well-defined bright circular ring emerges around the center. However, the intensity peak decreases with increasing $\omega$, while the radius of the bright ring expands.  Finally, we analyzed the effect of temperature on the holographic image. By examining the parameter $z_h$, which is the inverse of the event horizon radius $r_h$, we observed subtle variations in the size and brightness distribution of the rings. Since temperature is inversely proportional to $z_h$, higher values of $z_h$ correspond to lower temperatures. As $z_h$ increases, the peak brightness of the rings becomes more pronounced, meaning that the intensity of the holographic image decreases with increasing temperature.\\

Finally,  we analyzed the relationship between the incident angle of photons and the observed angle of the Einstein ring using numerical methods. By studying the radius of the Einstein ring for different values of the Gauss-Bonnet parameter $\alpha$, we demonstrated that the results obtained through wave optics in holography closely match those derived from geometric optics. Our findings showed that the Einstein radius computed from the response function aligns well with the geodesic predictions, with deviations limited to within three percent. This agreement validates the holographic approach for studying black hole images. \\

In conclusion, our study provides a comprehensive analysis of the holographic optical features of an AdS black hole with higher-derivative corrections and a string cloud. As a future direction, it would be interesting to extend this analysis to different black hole systems and explore how observational constraints on black hole parameters influence the resulting holographic images.

\section{Acknowledgments}
BH would like to thank DST-INSPIRE, Ministry of Science and Technology fellowship program, Govt. of India for awarding the DST/INSPIRE Fellowship[IF220255] for financial support. 	

	\bibliographystyle{apsrev}
	
\end{document}